\documentclass[sigconf,screen=true,bookmarks=false]{acmart}
\AtBeginDocument{%
  }
\usepackage[utf8]{inputenc} %
\usepackage[T1]{fontenc}    %

\usepackage{pifont}
\usepackage{xcolor}         %
\usepackage{booktabs}
\usepackage{hyperref}
\hypersetup{colorlinks,
    linkcolor=red,citecolor=citecolor,urlcolor=blue}
\usepackage{algorithm}
\usepackage{graphicx}
\usepackage{threeparttable}
\usepackage{multirow}
\usepackage{amsmath}
\usepackage{bm}
\usepackage{bbm}
\usepackage{amsfonts}
\usepackage{amsthm}
\usepackage[noend]{algpseudocode}
\usepackage{enumitem} %
\usepackage[subrefformat=parens,labelformat=parens]{subfig}
\usepackage{colortbl}
\usepackage[normalem]{ulem}
\usepackage{soul}
\usepackage{graphicx} %
\usepackage{booktabs}
\usepackage{algpseudocode}
\algtext*{EndIf}

\usepackage{wrapfig}

\newenvironment{problem}[1][Problem]
  {\begin{trivlist} \item[\hskip \labelsep {\bfseries #1.}]}
  {\end{trivlist}}

\usepackage{xspace}

\definecolor{citecolor}{RGB}{34,139,34}
\definecolor{mydarkblue}{rgb}{0,0.08,1}
\definecolor{mydarkgreen}{rgb}{0.02,0.6,0.02}
\definecolor{mydarkred}{rgb}{0.8,0.02,0.02}
\definecolor{mydarkorange}{rgb}{0.40,0.2,0.02}
\definecolor{mypurple}{RGB}{111,0,255}
\definecolor{myred}{rgb}{1.0,0.0,0.0}
\definecolor{mygold}{rgb}{0.75,0.6,0.12}
\definecolor{myblue}{rgb}{0,0.2,0.8}
\definecolor{mydarkgray}{rgb}{0.,0.2,0.2}

\definecolor{lightred}{RGB}{255,235,235}
\definecolor{lightgreen}{RGB}{235,255,235}
\definecolor{lightblue}{RGB}{235,235,255}
\definecolor{lightcyan}{RGB}{235,255,255}
\definecolor{lightmagenta}{RGB}{255,235,255}
\definecolor{lightyellow}{RGB}{255,255,235}

\definecolor{qxkcolor}{RGB}{215,235,255}
\definecolor{softmaxcolor}{RGB}{230,235,255}
\definecolor{probxvcolor}{RGB}{255,255,235}

\definecolor{topkcolor}{RGB}{255,235,235}
\definecolor{zecolor}{RGB}{255,255,235}
\definecolor{dynacolor}{RGB}{235,255,255}

\definecolor{reviewcolor}{RGB}{0,0,200}

\theoremstyle{plain}

\theoremstyle{definition}

\algdef{SE}[DOWHILE]{Do}{doWhile}{\algorithmicdo}[1]{\algorithmicwhile\ #1}%

\newcommand{\squishlist}{
 \begin{list}{$\bullet$}
  { \setlength{\itemsep}{0pt}
     \setlength{\parsep}{3pt}
     \setlength{\topsep}{3pt}
     \setlength{\partopsep}{0pt}
     \setlength{\leftmargin}{1.5em}
     \setlength{\labelwidth}{1em}
     \setlength{\labelsep}{0.5em} } }

\newcommand{\squishend}{
  \end{list}  }

\begin{document}

\settopmatter{printacmref=false} %
\renewcommand\footnotetextcopyrightpermission[1]{} %
\pagestyle{plain} %

\title{Photonics-Aware Planning-Guided Automated Electrical Routing for Large-Scale Active Photonic Integrated Circuits
}

\author
{
Hongjian Zhou$^1$, Haoyu Yang$^2$, Nicholas Gangi$^3$, Bowen Liu$^3$, Meng Zhang$^3$\\ Haoxing Ren$^2$, Xu Wang$^4$, Rena Huang$^3$, Jiaqi Gu$^{1\dagger}$\\
$^1$Arizona State University, $^2$NVIDIA, $^3$Rensselaer Polytechnic Institute, $^4$Cadence\\
{\small $\dagger$\emph{jiaqigu@asu.edu}}
}

\begin{abstract}
\label{abstract}
The rising demand for AI training and inference, as well as scientific computing, combined with stringent latency and energy budgets, is driving the adoption of integrated photonics for computing, sensing, and communications.
As active photonic integrated circuits (PICs) scale in device count and functional heterogeneity, physical implementation by manual scripting and ad-hoc edits is no longer tenable.
This creates an immediate need for an electronic–photonic design automation (EPDA) stack in which physical design automation is a core capability.
However, there is currently no end-to-end fully automated routing flow that coordinates photonic waveguides and on-chip metal interconnect.
Critically, available digital VLSI and analog/custom routers are not directly applicable to PIC metal routing due to a lack of customization to handle constraints induced by photonic devices and waveguides.
We present, to our knowledge, the first end-to-end routing framework for large-scale active PICs that jointly addresses waveguides and metal wires within a unified flow.
We introduce a physically-aware global planner that generates congestion- and crossing-aware routing guides while explicitly accounting for the placement of photonic components and waveguides.
We further propose a sequence-consistent track assignment and a soft guidance-assisted detailed routing to speed up the routing process with significantly optimized routability and via usage.
Evaluated on various large PIC designs, our router delivers fast, high-quality active PIC routing solutions with fewer vias, lower congestion, and competitive runtime relative to manual and existing VLSI router baselines; on average it reduce via count by $\sim$99\%, user-specified design rule violation by $\sim$98\%, and runtime by $17\times$, establishing a practical foundation for EPDA at system scale.

\end{abstract}

\maketitle
\vspace{-5pt}
\section{Introduction}
\label{sec:Introduction}

Photonic circuits offer intrinsic advantages, including low transmission loss and wide bandwidth that enable new classes of systems ranging from optical AI accelerators~\cite{NP_Nature2025_Hua} to chiplet-scale interconnects and on-chip sensing arrays~\cite{NP_CICC2024_Wang}.
By co-integrating optical and electrical components, electronic–photonic integrated circuits (EPICs) combine high-speed photonic communication with energy-efficient electronic control. 
This hybrid system enables new capabilities across domains: low-latency AI hardware, terabit-scale inter-chip links, and near-sensor signal processing, all within a compact form factor.
Fueled by advances in photonic device design and growing support from foundry process design kits (PDKs), EPICs are now being fabricated via multi-project wafers (MPWs)~\cite{NP_helkey2017introducing} and adopted by leading industry and academic efforts.
\begin{figure}
    \centering
    \includegraphics[width=0.99\columnwidth]{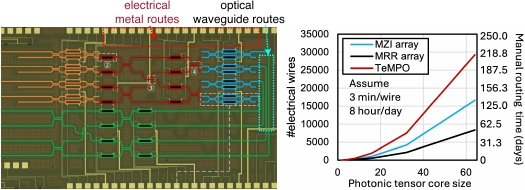}
    \vspace{-9pt}
    \caption{Electrical and optical routing in active PICs~\cite{NP_ACSPhotonics2022_Feng,NP_NATURE2017_Shen, NP_SciRep2017_Tait, zhang2024tempo}. Manually routing a large number of electrical metal wires consumes days to weeks.}
    \label{fig:Motivation}
     \vspace{-16pt}
\end{figure}

However, as EPICs scale to thousands of components, wires, and millimeter-scale routing paths, the physical design flow has become a dominant bottleneck.
Current methodologies remain largely schematic-driven and manual: engineers hand-place devices, script waveguide paths to meet curvature and phase constraints, and manually route electrical metal to pads with careful attention to port alignment, via insertion, layer assignment, and thermal/optical interference.
Commercial tools such as Synopsys OptoCompiler, Cadence Virtuoso, and GDSFactory provide schematic-driven layout environments and device libraries, but those semi-automated approaches still require extensive designer intervention for placement, routing, and layout closure.
As a result, layout closure remains time-consuming, error-prone, and non-scalable for large EPICs.

Recent research has addressed parts of this problem, especially in waveguide-centric physical design. Tools such as PROTON~\cite{boos2013proton}, PLATON~\cite{von2016platon}, and PlanarONoC~\cite{chuang2018planaronoc} automate global waveguide routing using loss-aware pathfinding and grid-based methods. 
Others, such as Apollo~\cite{NP_ICCAD2025_Zhou} and LiDAR~\cite{PD_ISPD2025_Zhou}, have demonstrated scalable placement and curvy-aware waveguide routing with design-rule-compliant GDS generation. 
These tools represent real progress toward fabrication-ready electronic–photonic design automation (EPDA), particularly for passive photonic circuits.

Yet a critical gap remains: electrical routing automation for active PICs is not well supported, despite being essential to functionality and layout closure.
In contrast to passive photonic circuits, active PICs, especially those computing circuits, integrate dense arrays of modulators, resonators, photodetectors, and phase shifters, requiring dense electrical interconnects for biasing, signaling, ground, or readout. As shown in Fig.~\ref{fig:Motivation},
designers must carefully plan metal wires and I/O pad connections for active devices, find feasible routing solutions in limited metal layer resources, while ensuring compliance with both electronic design rules and photonic layout constraints. These steps account for a significant fraction of the total layout effort and are not scalable.

PIC electrical routing presents \uline{three fundamental challenges} not addressed by conventional VLSI or analog routing tools:
\ding{202}~\textbf{Near-planar routing with high wire density and length}. Most photonic processes offer a few metal layers, e.g., 2 layers in AIM Photonics, and 8 layers in GlobalFoundries, with no layer-based routing directionality, which requires crossing-minimized wire planning and avoiding excessive via insertion.
At the same time, large EPICs feature thousands of nets, many spanning millimeters across the die, leading to severe congestion and runtime challenges.
\ding{203}~\textbf{Electrical–optical interaction constraints.}~ Routing must avoid long, co-parallel overlap between metal and waveguides to limit thermal coupling and optical loss. Additionally, designers frequently impose custom isolation margins, beyond minimum spacing design rules, near critical photonic structures such as modulators and resonators.
\ding{204}~\textbf{Packaging-aware pad assignment and breakout.}~
PICs must respect fixed electrical and optical I/O pad locations, often driven by fiber array pitch or wirebonding constraints. 
This introduces geometric constraints on I/O breakout and enforces routing order preservation along chip edges.

Despite its critical role, electrical routing in EPICs remains a manual, fragmented process, handled independently from waveguide routing and absent from any integrated EPDA flow.

\begin{figure}
    \centering
    \includegraphics[width=0.99\columnwidth]{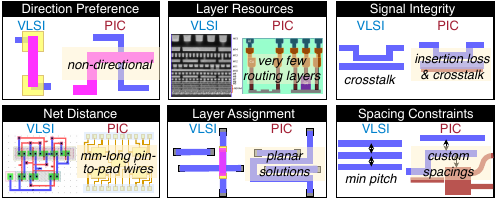}
    \vspace{-10pt}
    \caption{Compare electrical routing properties in VLSI circuits and active PICs.}
    \label{fig:Compare}
    \vspace{-15pt}
\end{figure}

To address this gap, we propose the first end-to-end framework for large-scale active photonic integrated circuits that integrates electrical routing and waveguide routing into a unified, physically-aware flow.
We introduce a global electrical routing planner that is congestion- and DRC-aware, explicitly accounts for the placement of photonic components and waveguides, and minimizes crossing and via usage. 
It further performs waveguide-aware track assignment and applies a guidance-driven detailed router that preserves route quality under tight spacing and co-propagation constraints with significantly faster runtime than prior un-guided detailed router.
Our key contributions are as follows:
\begin{itemize}
\vspace{-4pt}
    \item \textbf{End-to-End Automated Routing Framework for Active PICs} -- We develop the first automated flow that jointly routes electrical metal and photonic waveguides, targeting manufacturable, layout-rule-compliant PIC physical designs.
    \item \textbf{Physically-Aware Global Planning} -- We introduce a photonics- and crossing-aware electrical global routing that minimizes wire crossings, metal–waveguide interaction, and via usage, while honoring user-specified design constraints.
    \item \textbf{Soft Guidance-Assisted Detailed Routing} -- Our detailed router follows global guides and enforces spacing constraints with optical components, achieving high route quality and runtime efficiency in the near-planar regime.
    \item We extensively evaluate our router on large-scale active PIC benchmarks, and our framework achieves lower congestion, fewer vias and crossings, and competitive runtime relative to manual design and adapted VLSI routers, marking a significant step toward scalable, fabrication-ready EPDA.
    \emph{We will open-source codes and our active PIC benchmark suites.}
\end{itemize}

\vspace{-10pt}
\section{Preliminaries}
\label{sec:Preliminaries}

Routing in PICs comprises two domains: waveguide routing and electrical routing. Waveguide routing interconnects component optical ports, whereas electrical routing connects the electrical pins of active devices (e.g., the control wires that drive Mach-Zehnder modulators). 
The principal challenges in waveguide routing lie in enforcing curvilinear structures and automatically crossing insertion while minimizing insertion loss, which has been addressed in work~\cite{PD_ISPD2025_Zhou}.
In this section, we primarily focus on electrical routing for PICs by delineating its distinctions from conventional routing formulations and unique challenges.

\vspace{-5pt}
\subsection{Understanding the Challenges of PIC Electrical Routing 
} 

Automated routing in active PICs involves two physically distinct but tightly coupled domains: \textbf{waveguide routing}, which connects optical ports through curvature- and phase-constrained paths in the lowest silicon layer, and \textbf{electrical routing}, which establishes electrical connectivity for control, biasing, and readout of active components. 
While recent work has focused on automating waveguide routing, electrical routing in active PICs remains largely manual.

Beyond the basic design rules, minimum wire spacing, wire width, and metal area, several factors distinguish PIC electrical routing from VLSI and analog IC routing, as shown in Fig.~\ref{fig:Compare}

\noindent\textbf{Limited Routing Layers and No Direction Preference}.~ Most foundry PIC platforms expose only a small number of metal layers—typically two in AIM Photonics and up to eight in GlobalFoundries. 
In practice, designers often choose to restrict usage to fewer layers to reduce fabrication cost (e.g., fewer masks) and to minimize via-induced parasitics such as resistance, capacitance, and thermal hotspots. 
Unlike digital VLSI, where metal layers are organized with orthogonal directionality (e.g., horizontal M2, vertical M3), PIC electrical routing lacks this structure. 
Electrical nets in active PICs are often used not for internal logic connections but for I/O, with device pins fanned out toward peripheral pads for wirebonding. 
These factors result in a \underline{near-planar} routing regime: \emph{limited layers, long nets, dense component regions, and unconstrained directions}. 
Routing thousands of such nets within this regime demands careful congestion management, via minimization, and global planning across the entire die—challenges that traditional VLSI routers are ill-equipped to handle.

\noindent\textbf{Electrical-Photonic Interaction Constraints}.~
While metal wires can legally pass over photonic devices if no electrical conflict exists (i.e., short circuits), sustained co-propagation above waveguides and excessive overlap with photonic components are often discouraged.
Metal structures near optical paths can introduce absorption, reflection, and thermal coupling, which degrade insertion loss and tuning efficiency. 
To mitigate this, designers prioritize avoiding the lower-layer metal wires from overlapping with:
active photonic devices (e.g., heaters, phase shifters, modulators); passive components (e.g., directional couplers, Y-branches), and previously routed waveguides with long coupling length.
These components act as \underline{(soft) blockages}, reducing available routing area and \emph{exacerbating congestion in already near-planar layout}.

\noindent\textbf{Non-Standard, User-Specified Spacing Rules}.~
Beyond foundry-defined design rules, minimum spacing (e.g., 1 $\mu m$), width, and metal density, PIC designers usually impose \uline{custom, often aggressive, spacing constraints} near sensitive photonic elements, sometimes exceeding 10 $\mu m$, to suppress electrical crosstalk, thermal hotspots, and layout-induced parasitics.
These user-specified constraints are circuit- and platform-specific, and are not captured by a generic design rule checker (DRC) in traditional EDA tools.

\noindent\textbf{Long Wires and Routing Scalability}.~
PICs frequently require millimeter-scale metal wires for connecting heaters, modulators, and detectors across large dies.
Routing such \uline{long nets over a fine grid} (as used in VLSI) becomes runtime expensive due to grid explosion, while coarse grids suffer from poor resolution and frequent failure in congested regions.
This creates a \textbf{scalability bottleneck}: fast routing engines must balance grid granularity with physical realism, particularly under complex component-aware constraints.

Together, these challenges make PIC electrical routing incompatible with standard VLSI/analog routers.
In this work, we address these challenges by building a co-routing framework from the ground up, tailored to the unique constraints of active PICs.

\vspace{-5pt}
\subsection{Related Work}
Although prior work on PIC electrical routing is lacking, related routing problems have been extensively explored in adjacent domains such as PCB/MCM routing~\cite{PD_lin2021complete, PD_lin2021unified, Routing_DAC23_Liu} and packaging routing~\cite{PD_fang2005routing, PD_fang2007network, PD_yu2019flip}. 
For example, in redistribution-layer (RDL) routing, the number of usable layers is typically very limited~\cite{PD_fang2008area}, so obtaining a (near-)planar graph solution is often a design objective. 
In peripheral I/O routing~\cite{PD_fang2005routing, PD_fang2007network}, external pads must be routed to internal bumps, which introduces a pad-assignment problem~\cite{PD_yu2019flip}. 
In this setting, pin distributions in packaging are relatively regular, and there are typically no large blockages. 
Consequently, many works~\cite{PD_lee2012obstacle,PD_fang2005routing,PD_fang2007network} adopt minimum-cost maximum-flow (MCMF) formulations; the underlying bipartite/grid graph is inherently planar in the absence of obstacles, so the solution yields crossing-free global routing paths for each net.
The widely-studied PCB routing problem is to route all the connections between the BGA of chip packages while satisfying various physical constraints. Besides, the use of vias is often restricted, and thus the routing becomes planar. Previous studies categorized the PCB routing problem into two types: (1) the escape routing problem~\cite{PD_lei2015optimizing, PD_kong2010optimal} and (2) the area routing problem~\cite{PD_yan2010recent, PD_yan2011obstacle}. The escape routing problem is to route from the pads of BGAs to their array boundaries. The works~\cite{PD_ozdal2006algorithms,PD_yan2018layer} carry out layer assignment either following pin escape or concurrently with it, thereby limiting the number of vias employed during routing.
The area routing problem is to connect the previously escaped routes of BGAs. At this stage, most works focus on length matching~\cite{PD_yan2009bsg, PD_yan2011obstacle}.
Despite these similarities (planarity, peripheral I/O), these methods cannot be directly applied to PIC electrical routing: the irregular distribution of photonic components prevents the straightforward construction of a crossing-free graph, and the routing must additionally account for routed waveguides and user-defined spacing rules.

\vspace{-5pt}
\subsection{Problem Formulation}
\label{subsec:problem_formulation}
In this paper, we present (to the best of our knowledge) the first automated flow that performs both waveguide and electrical routing for PICs. Concretely, we decouple the two subproblems: waveguide routing is first carried out using LiDAR~\cite{PD_ISPD2025_Zhou}, followed by electrical routing. Our technical contributions primarily concern the electrical routing stage. We now formally define the PIC electrical routing problem as follows:
\begin{problem}[PIC Electrical Routing]
Given a set of electrical nets $N_e$, a set of placed photonic components, and the user-specified preferences, generate a routing solution for each $n_i\in N_e$, so that there are no design rule violations and the specified preferences are satisfied.
\end{problem}

\section{Automated Active PIC Electrical Routing}
In this section, we present our metal routing algorithm for PIC. First, we give an overview of our proposed algorithm. Then, we detail the methods used in each stage of the algorithm.

\vspace{-5pt}
\subsection{Algorithm Overview}
\begin{figure}
    \centering
    \includegraphics[width=0.55\columnwidth]{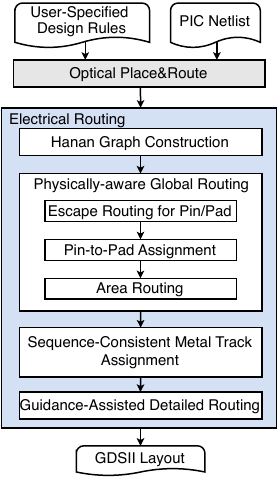}
    \vspace{-10pt}
    \caption{Algorithm flow of the proposed active PIC electrical routing framework. 
    }
    \label{fig:Overview}
    \vspace{-13pt}
\end{figure}

Figure~\ref{fig:Overview} summarizes our routing framework. 
Given an input netlist consisting of optical and electrical nets, we first generate a manual placement and perform waveguide routing with LiDAR~\cite{PD_ISPD2025_Zhou} to obtain the routed waveguide geometry. 
We then perform physically-aware global planning for the electrical nets, including pad assignment and the construction of routing guidance. The guidances are generated by two steps: (1) crossing-aware A* routing on Hanan grids that treats photonic components as blockages to obtain a (near-)planar solution, and (2) track assignment that respects the routed waveguides and user-specified design rules. Finally, these guidance cues are used to steer detailed routing.
We next detail each component of the electrical routing stage.

\begin{figure}
    \centering
    \includegraphics[width=1\columnwidth]{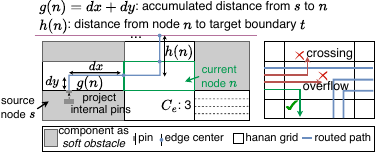}
    \vspace{-10pt}
    \caption{(a) Hanan routing grid construction and $g(n)$ and $h(n)$ cost calculation. (b) crossing check and capacity check.
    }
    \label{fig:Grid}
    \vspace{-5pt}
\end{figure}

\vspace{-5pt}
\subsection{Hanan Routing Grid Initialization}
While direct A*-based detailed routing is commonly used in analog and custom design, it becomes \textbf{intractable and error-prone when applied directly to large PIC layouts}. 
Without global planning, A$^\ast$ often produces tangled paths, excessive crossings, or frequent ripup\&reroute and routing failures in congested regions.
To address these limitations, we adopt a physically-aware planning-guided routing flow
, in which a global planning stage first computes a photonic-aware, capacity-constrained routing guide across multiple metal layers. This guide informs subsequent track assignment and detailed routing, improving routability and scalability.

To make metal routing photonic-aware and enable obstacle avoidance, we construct a customized Hanan grid over the layout, which encodes the layout geometry, metal-layer capacities, and photonic component locations, as described below.

The grid is initialized using the set of electrical pin coordinates and photonic component bounding boxes, which treats photonic components as \emph{soft obstacles}. 
Pins located inside the component are first projected to the nearest grid boundary (along the shortest distance to the component boundary) before routing, as shown in Fig.~\ref{fig:Grid}.
Unlike traditional Hanan grids, which embed nets as axis-aligned paths on graph edges, our grid supports \textbf{cell-based routing}, where paths \emph{traverse between cell centers}, crossing boundaries orthogonally.
For each Hanan grid edge $e$ on layer \(\ell\), we compute a \textbf{capacity}
\(c_e=\big\lfloor \mathrm{length}(e)/p_\ell \big\rfloor\),
where \(p_\ell\) is the minimum wire pitch of layer \(\ell\), specified by the process design kit (PDK). 
These capacities define the maximum number of parallel wire tracks that may legally traverse each edge, enabling congestion-aware multi-layer global metal planning. While the Hanan grid is defined in three dimensions, layers are decoupled; that is, grid graphs on distinct layers are independent, and no inter-layer edges are present.

\vspace{-5pt}
\subsection{Physically-Aware Global Routing}
In the global routing stage, our goal is to generate crossing-aware routing guides.
We propose a \emph{crossing-aware A$^*$ planning engine} that enforces consistent net ordering across routing boundaries to minimize topological crossings.
While conventional A$^*$ search is unaware of multi-net interactions and often produces intersecting routes, our method guarantees planarity by maintaining a global net ordering on each grid edge and pruning paths that would violate this ordering, inspired by prior work~\cite{PD_chung2023any}.

\noindent\textbf{Crossing-Aware A* Planning Engine}.~
The standard formulation of the node cost for the current node $n$ in the A$^\ast$ search is given by: $f(n) = g(n) + h(n)$, where $g(n)$ denotes the accumulated cost from the source node $s$ to the current node $n$, and $h(n)$ is the heuristic estimate of the cost from node $n$ to the target node $t$. 
As illustrated in Fig.~\ref{fig:Grid}, when \uline{computing $g(n)$}, we use the \emph{actual pin location} at the source node, whereas for all other nodes we \emph{measure distances with respect to the center point of the incident edge} since the actual tracks have not been assigned.
The \uline{heuristic cost $h(n)$} is defined as the distance from the current node $n$ to the \emph{target bounding box boundary $t$}.
Driven by the above-defined cost, our proposed crossing-aware A$^\ast$ search engine \textbf{tracks the net order that passes through each node edge in the routing grid} with a \textbf{net sequence list}: an ordered set of nets that traverse that edge.
If two net path orders on the grid cell boundary become inconsistent, e.g., net1 and net2's orders are swapped on the incident and outgoing grid edges, those two electrical routing paths will cross each other and introduce a via. Figure~\ref{fig:Grid} gives the example of crossing check and capacity check during routing. 
Therefore, during A$^\ast$ search for a given net, we try to find a planar (non-crossing) routing solution by maintaining a \textbf{consistent net order constraint at each grid node edge}, thereby generating a non-crossing topology.

While this net-sequence-based constraint guarantees the generation of a non-crossing guide for each individual net, the procedure is \textbf{inherently sequential and order-dependent}, and thus does not guarantee global optimality.
Early-routed nets may consume critical routing corridors, block escape routes, and make later nets violate planarity, thus forcing via insertions and increasing metal layer usage.
When no planar solution exists on the current layer, we permit \textbf{routing layer switching at endpoints only}, i.e., a net may route on a different layer from source to target, but \emph{mid-path via insertions} are disallowed to preserve simplicity.
If no feasible guides exist in any routing layers, the planning of this net will be skipped with no solution found, and the procedure for the rest of the nets will continue.

\noindent\textbf{Multi-Stage Planning Flow: Pin/Pad Escape Routing $\rightarrow$ Pin-to-Pad Assignment $\rightarrow$ Area Routing}.~
Since the metal wire crossings are mostly caused by \emph{sub-optimal net-ordering and pin-to-pad assignment}, we propose to manage this complicated problem by decomposing global planning into \underline{three structured stages}: (1) \emph{escape routing}, (2) \emph{pin-to-pad assignment}, and (3) \emph{area routing}.

\begin{figure}
    \centering
    \includegraphics[width=0.85\columnwidth]{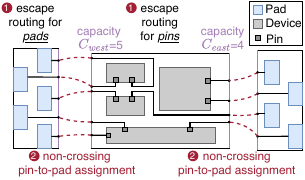}
    \caption{Illustration of \ding{202}~pin/pad escape routing and \ding{203}~pin-to-pad assignment. 
    Escape net based on the distance to the boundary while satisfying the boundary edge capacity. Then assign pads based on the net order on the boundary to ensure a non-crossing assignment result.
    }
    \label{fig:Escape}
    \vspace{-10pt}
\end{figure}

\subsubsection{Pin/Pad Escape Routing}
Rather than routing each electrical pin directly to its target pad, we first \textbf{route all pins to the boundary} of a common axis-aligned bounding box enclosing all active components, as shown in Fig.~\ref{fig:Escape}\ding{202}.
This single, shared target (box boundary) avoids the net permutation chaos and thus significantly reduces path crossing risks, regularizes breakout topology, and simplifies pad assignment.

First, we determine the processing order for each pin.
For a pin \(p\), we compute its distance to the box boundary as the minimum of its distances to the four sides and process pins in non-descending order of this distance (nearer pins receive higher priority).

To avoid overloading any one side, we assign a \uline{side capacity to each edge of the bounding box}, equal to the number of I/O pads available on that side.
During escape routing, pins are assigned to bounding box edges under this capacity constraint, preventing escape congestion due to demand-resource imbalance and reducing downstream failure in the later area routing stage. And the A$^\ast$ heuristic cost $h(n)$ is set to the Manhattan distance from the current node to the boundary.
We apply a similar \emph{escape procedure to the pad arrays}, especially when they contain multiple rows or staggered placement, which \underline{ensures intra-pad-array planarity} before completing routes from the box boundary to the assigned pads.

By performing escape routing on both the device pins and the I/O pads, we effectively \underline{build planar routing "bridges" from both ends}.
This strategy transforms the original complex many-to-many routing problem into a one-to-one mapping-and-route problem, and greatly simplifies the final connection stage.

\vspace{-5pt}
\subsubsection{Pin-to-Pad Assignment and Area Routing}
A key advantage of performing escape routing first is that pin-to-pad assignment can be decided during routing. Otherwise, one must precompute a pin–to–pad mapping (e.g., via linear programming) prior to routing; without awareness of photonic components and keep-outs, such preassignment often fails to preserve planarity and complicates downstream routing. 

After escaping all pad arrays and device pins to a common bounding-box boundary, we perform pin-to-pad assignment as illustrated in Fig.~\ref{fig:Escape}\ding{203}.
The procedure is as follows: for each side of the boundary, we traverse the escaped pins following their orders on the boundary and assign them to pads on the same side in the corresponding order, respecting side capacities. 
Because the region outside the bounding box contains no active photonic components, the subsequent area routing from the boundary to the assigned pads can be carried out without crossings. 
For the area-routing stage, we reuse the crossing-aware A$^\ast$ search engine; the only difference from pin/pad escape routing is that the source–target paths are searched from the bounding-box boundary to the pad-array boundary.

\begin{figure}
    \centering
    \includegraphics[width=1\columnwidth]{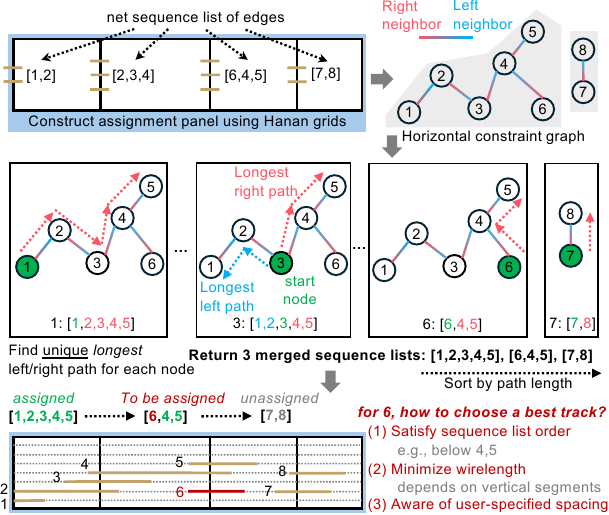}
    \vspace{-14pt}
    \caption{Overall track assignment procedure. An assignment panel and constraint graph are constructed using the Hanan grids and the net sequence list of each edge. Then, assign the segment using the merged sequence list according to wirelength/spacing considerations.
    }
    \label{fig:Graph}
\end{figure}

\vspace{-5pt}
\subsection{Sequence-Consistent Track Assignment
}

After global routing, each net has a \textbf{sequence of Hanan grid nodes} that it traverses.
The next critical step is to determine \textbf{specific track placement} within each node, while respecting the \emph{net ordering} established on each edge during global routing.
This stage produces a \textbf{fine-grained routing guidance} that ensures high routability, reduces crossings, and minimizes parallel metal–waveguide overlap.
This procedure is outlined in Alg.~\ref{alg:track_assignment}.

\noindent\textbf{Wire Segment Decomposition}.~
We first decompose all routed paths into \uline{horizontal and vertical segments} for each node they traversed.
Track assignment proceeds in \uline{two passes}:
(1) Horizontal segments are assigned tracks, and the lengths of their vertical connections are updated;
(2) Then, vertical segments are assigned, and the corresponding horizontal lengths are updated.

\begin{algorithm}[t]
\caption{Track Assignment for Horizontal Segments}
\label{alg:track_assignment}
\begin{algorithmic}[1]
\Require Segments $\mathcal{S}$, net-sequence lists $L$, user-sepcified rules $\mathcal{R}$
\Ensure Per-net detailed routing guidance $\mathcal{O}$
\For{each $s \in \mathcal{S}$}
  \If{$s.\mathrm{assigned}$} \State \textbf{continue} \EndIf
  \State $\mathcal{P} \gets \textsc{BuildPanel}(s,\mathcal{S})$
  \Comment{using adjacent Hanan grids}
  \State $G=(V,E) \gets \textsc{BuildConflictGraph}(\mathcal{P})$
  \State $\mathcal{M} \gets \{\,\textsc{MergedSequence}(v,L, G)\mid v\in V\,\}$
  \State $\mathcal{M} \gets \textsc{SortSequencesByDepth}(\mathcal{M})$ \Comment{deepest-first}
  \For{each merged sequence $m \in \mathcal{M}$}
    \For{each $u \in \textsc{OutsideToInsideOrder}(m,\mathcal{P})$}
      \If{$u.\mathrm{assigned}$} \State \textbf{continue} \EndIf
      \State \textsc{AssignTrack}$(u,\mathcal{P},\mathcal{R})$
      \State \textsc{UpdateVerticalSegment}$(u)$
      \State $u.\mathrm{assigned} \gets \texttt{true}$
    \EndFor
  \EndFor
\EndFor
\State $\mathcal{O} \gets \textsc{StitchGuidelines}(\mathcal{S})$
\end{algorithmic}
\end{algorithm}

\noindent\textbf{Panel-Based Joint Track Assignment}.~
Rather than \emph{locally} assigning tracks node by node, we introduce a \textbf{novel panel-based assignment} strategy with a \emph{larger planning view}.
A panel is a \emph{window of contiguous grid nodes along a row or column}, containing all segments that may interact or compete for tracks.
Within each panel, we leverage the net sequence information from global routing to perform \emph{joint} track assignment.
Specifically, we identify all segments that lie on the same row as the segment to be assigned, take the union of the grid nodes they occupy, and construct a panel for joint track assignment. 

\ding{202}~\emph{\uline{Constraint Graph Construction}}: 
For all segments within the panel, we construct a constraint graph $G=(V,E)$, where each vertex represents a segment.
As shown in Fig.~\ref{fig:Graph}, each vertex is connected to left neighbor nodes via blue edges and connected to right neighbor nodes via red edges.
In the horizontal constraint graph example in the figure, given the net sequence [2,3,4] and [6,4,5], node 4 has two left nodes 3 and 6, and one right node 5.

\ding{203}~\emph{\uline{Merging Net-Sequences}}:
To identify the maximal mergeable net-sequence blocks, we find the longest left path (defined as the longest path that only contains edges that point to left neighbors) and longest right path for each node simply using a recursive algorithm.
Then it forms the longest mergeable net-sequence by concatenating the longest left path, the start node itself, and the longest right path, e.g., for start node=3, we have [1,2] | [3] | [4,5]$\rightarrow$[1,2,3,4,5].
After deduplication, we find all \textbf{unique longest mergeable net-sequences}.
The merged sequences will be ranked by depth, with deeper ones assigned to tracks first.
This \textbf{deepest-first ordering} ensures that subsequent inner segments have sufficient available tracks, reducing conflicts and resource contention.

\ding{204}~\emph{\uline{Wirelength-minimized Spacing-honored Track Assignment}}:
In each \emph{panel}, track positions are discretized using the \emph{layer-specific metal wire pitch}.
For each segment in a merged net-sequence list, we assign tracks according to the following \uline{\textbf{rules}}:
\textbf{(1)}~\textbf{Order Constraint}: We select available tracks that satisfy the sequence list order, e.g., in the example given in the figure, only the 3 unoccupied tracks below the assigned segment 4 can be selected as candidate tracks for segment 6.
\textbf{(2)}~\textbf{Wirelength Minimization}: 
Different track assignment can potentially induce different wirelength, especially when the two connecting segments of the interested segment are on the same side.
If both neighboring segments of the same net are pointing in the same direction, we will bias the track toward that direction to minimize wirelength and select the nearest available track.
If neighbors lie on opposite sides, any track yields equivalent Manhattan length.
\textbf{(3)}~\textbf{Ouside-In Assignment}:
Tracks are assigned from the outermost segments of the panel inward. 
This ordering simplifies resource allocation and prevents later conflicts.
\textbf{(4)}~\textbf{Pin Access and Spacing Constraints}:
Segments that are directly connected to pins must lie within the pin’s access window.
Meanwhile, user-preferred wire spacing is enforced between segments; if insufficient space remains, spacing is reduced to the maximum feasible value.
\textbf{(5)}~\textbf{Optical Waveguide Interaction Check}:
After assignment, we check each segment’s parallel overlap with pre-routed waveguides. If the overlap exceeds a threshold \(L_{\mathrm{th}}\), the segment is shifted by at most one track to reduce unexpected metal-induced waveguide insertion loss; otherwise, the assignment is accepted.

Finally, the assigned segments are \emph{stitched into continuous routing guidelines for each net}, which serve as input for the detailed routing stage, ensuring significantly faster routing and higher planar solution quality that respects both electrical and optical constraints.

\begin{figure}
    \centering
    \includegraphics[width=0.78\columnwidth]{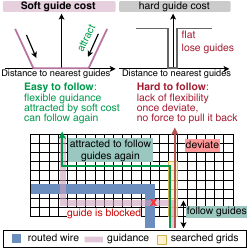}
    \vspace{-5pt}
    \caption{Proposed soft guidance provides a flexible mechanism to attract the searching node to follow the guidance even with blocked guidance and temporary search path deviation, and thus can accelerate the routing process.
    }
    \label{fig:Guide}
    \vspace{-10pt}
\end{figure}

\vspace{-5pt}
\subsection{Soft Guidance-Assisted Detailed Routing}
After track assignment, fine-grained detailed routing is required to generate DRC-compliant metal paths that satisfy all pin access, spacing, and photonic constraints.
While standard A$^\ast$-based detailed routers can perform grid-based routing with electrical DRC checks, they \textbf{struggle in the PIC context} as they (1) \uline{cannot handle electrical-photonic interactions}, as metal wires that overlap or run close to photonic components can degrade optical performance via absorption or thermal coupling; (2) \uline{cannot effectively follow global routing guides}: track guidelines may become blocked due to pin placement constraints or already routed wires. \textbf{Hard on-guideline} enforcement often causes the search to lose the intended path and drift away, resulting in unnecessary detours or routing failure, as illustrated in Fig.~\ref{fig:Guide}.

To address these issues, we extend an open-source analog detailed router, \texttt{Anaroute}~\cite{PD_IEEEDT2021_Chen}, with two key enhancements:
\textbf{(1)~Photonic Awareness}: We impose additional penalties on routing paths that overlap or are too close to photonic components and waveguides.
This ensures metal paths respect photonic layout constraints.
\textbf{(2)~Soft Guidance for Track Following}:
Instead of a hard “on-guideline” constraint, we propose a critical technique: \textbf{distance-based soft penalty} that increases as a wire deviates from the assigned guideline.
This encourages the routing search to stay close to the global routing guidance, while still being flexible, allowing detours in congested/blocked regions.
When a guideline segment is blocked by pin access or previously routed wires, the soft penalty allows the search to \textbf{temporarily deviate but then attracts the path back} toward the guideline once the blockage is cleared.
Figure~\ref{fig:Guide} illustrates this behavior: 
A hard-guideline search would cause the path to drift away and lose guidance, whereas soft guidance maintains path fidelity while providing flexibility, producing smoother, DRC-compliant routing that honors both electrical and photonic constraints.

\section{Evaluation Results}
\begin{table}
  \centering
  \caption{Active PIC benchmark information.
  }
  \label{tab:bench_stat}
  \vspace{-10pt}
  \resizebox{\columnwidth}{!}{
  \begin{tabular}{|c|c|c|c|c|}
    \hline
    Benchmark & \#Devices & \#Optical Nets & \#Electrical Nets & Die Size \\ \hline
    \texttt{Clements}\_8$\times$8~\cite{NP_Optica2018_Clements}                     &132  &87  &71  & $3600\times1800\ \mu\mathrm{m}^2$ \\
    \texttt{Clements}\_16$\times$16~\cite{NP_Optica2018_Clements}                   &456  &303  &271  & $7600\times 3300\ \mu\mathrm{m}^2$ \\
    \texttt{Clements}\_32$\times$32~\cite{NP_Optica2018_Clements}                   &1682  &1119  &1056  & $14300\times 6100\ \mu\mathrm{m}^2$ \\
    \texttt{ADEPT}\_8$\times$8~\cite{NP_DAC2022_Gu}                                 &232  &127  &127  & $4500\times2000\ \mu\mathrm{m}^2$ \\
    \texttt{ADEPT}\_16$\times$16~\cite{NP_DAC2022_Gu}                               &547  &319  &319  & $7400\times3600\ \mu\mathrm{m}^2$ \\
    \texttt{ADEPT}\_32$\times$32~\cite{NP_DAC2022_Gu}                               &1299  &767  &767  & $100000\times6000\ \mu\mathrm{m}^2$ \\
    \texttt{GWOR}\_8$\times$8~\cite{PD_tan2011generic}                              &118  &32   &85  & $3000\times 3000\ \mu\mathrm{m}^2$ \\
    \texttt{GWOR}\_10$\times$10~\cite{PD_tan2011generic}                            &190  &50   &160  & $5100\times5100\ \mu\mathrm{m}^2$ \\
    \hline
  \end{tabular}
  }
  \vspace{-13pt}
\end{table}

\begin{figure*}[!t]
  \centering
  \vspace{-10pt}
  \includegraphics[width=0.95\textwidth]{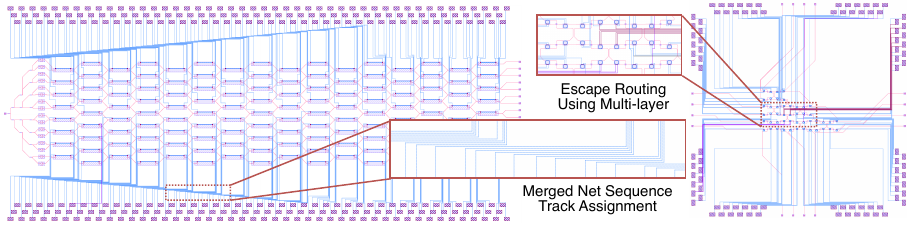}
  \vspace{-13pt}
  \caption{Visualization of our generated complete PIC layout (Clements$\_16\times 16$~\cite{NP_NATURE2017_Shen}, GWOR$\_8\times 8$~\cite{PD_tan2011generic}) with automatically routed optical waveguides and electrical connections.
  The zoomed-in regions highlight the planar routing solution from our multi-layer escape routing and net-sequence-based track assignment algorithms.}
  \label{fig:Layout}
  \vspace{-5pt}
\end{figure*}

\begin{figure}[t]
  \centering
  \includegraphics[width=\columnwidth]{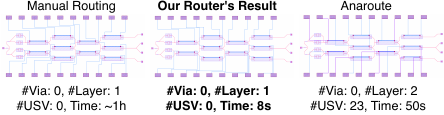}
  \vspace{-15pt}
  \caption{Comparison of manual and automated routing of (Clements$\_4\times 4$~\cite{NP_NATURE2017_Shen}). Our router achieves comparable results to manual routing while requiring substantially less time, while naive Anaroute fails to find a planar routing solution, and the routed metal wires cross the photonic components.}
  \label{fig:Manual}
  \vspace{-15pt}
\end{figure}

\begin{table*}[t]
\centering
\caption{ Comparisons of the used via number (\#Via), used routing layer (\#Layer), number of user-specified design rule violations (\#USV), runtime (s), and the total wirelength (WL ($mm$)).
$\downarrow$: lower is better.}
\vspace{-8pt}
\resizebox{\textwidth}{!}{
\begin{tabular}{cccccccccccccccc}
\hline
& \multicolumn{5}{c}{Anaroute~\cite{PD_xu2019magical} }         & \multicolumn{5}{c}{Anaroute$^\ast$~\cite{PD_xu2019magical}}                 & \multicolumn{5}{c}{\cellcolor[HTML]{D9D9D9}\textbf{Our Proposed Router}}                     \\
\cmidrule(lr){2-6} \cmidrule(lr){7-11} \cmidrule(lr){12-16}
\multirow{-2}{*}{Benchmark} & \#Via$\downarrow$ & \#Layer$\downarrow$ & \#USV$\downarrow$    & Runtime (s)$\downarrow$ & WL ($mm$)$\downarrow$       & \#Via$\downarrow$ & \#Layer$\downarrow$ & \#USV$\downarrow$    & Runtime (s)$\downarrow$ & WL ($mm$)$\downarrow$    & \#Via$\downarrow$ & \#Layer$\downarrow$ & \#USV$\downarrow$    & Runtime (s)$\downarrow$ & WL ($mm$)$\downarrow$   \\
\hline
Clements\_8$\times$8            &   1  & 2 & 105 & 45   & 42.99  & 2   & 3 & 23  & 114  & 45.86  & 0 & 1 & 2  & 33  & 46.94  \\
Clements\_16$\times$16          &  69 & 3 & 879 & 2383 & 340.97 & 117 & 3 & 370 & 2324 & 348.69 & 0 & 1 & 4  & 207 & 360.51 \\
Clements\_32$\times$32          & 159 & 3 & 2374 & 13235 & 2084.42 & 205 & 3 & 753 & 20093 & 2095.78 & 0 & 2 & 12 & 930 & 2124.23 \\
ADEPT\_8$\times$8               &   8  & 3 & 178 & 212  & 95.58  & 16  & 3 & 74  & 481  & 97.2   & 0 & 1 & 2  & 65  & 100.7  \\
ADEPT\_16$\times$16             &   54 & 3 & 674 & 1558 & 422.9  & 83  & 3 & 249 & 3184 & 425.98 & 0 & 1 & 22 & 245 & 440.56 \\
ADEPT\_32$\times$32             &   402 & 3 & 1603 & 10696 & 2319.2 & 519 & 3 & 853 & 17523 & 2325.31 & 0 & 2 & 37 & 871 & 2347.02 \\
GWOR\_8$\times$8                &   43 & 3 & 357 & 2494 & 181.8  & 64  & 3 & 248 & 3343 & 185.02 & 6 & 3 & 19 & 99  & 184.84 \\
GWOR\_10$\times$10              &   173 & 3 & 2904 & 17281 & 561.20 & 228 & 3 & 1347 & 24960 & 566.72 & 2 & 3 & 26 & 316 & 567.5 \\ 
\hline
Geo-mean                        & 114 & - & 1134 & 5988 & 756.13 & 154 & - & 490 & 9003 & 761.32 & 1 & - & 16 & 346 & 771.53 \\
ratio                           & 1.000 & - & 1.000 & 1.000 & 1.000   & 1.351 & -  & 0.432 & 1.504 & 1.007   & 0.008 & -  & 0.014 & 0.058 & 1.020 \\
\hline
\end{tabular}
}
\label{tab:result}
\vspace{-10pt}
\end{table*}

The proposed framework for PIC electrical routing and waveguide routing is implemented in Python and C++. The global routing is developed using Python, while the detailed routing is developed based on the router \texttt{Anaroute} in MAGICAL~\cite{PD_xu2019magical}. 
The waveguide routing is done by an open-source PIC waveguide router \texttt{LiDAR}~\cite{PD_ISPD2025_Zhou}. 
All evaluations are conducted on a Linux server with a 128-core AMD EPYC 7763 CPU and 512 GB memory.

\noindent\underline{\textbf{Benchmarks.}}~
We evaluate our router on large-scale active PIC benchmarks in computing and interconnect applications, derived from publicly available PIC designs~\cite{PD_ISPD2025_Zhou}. 
For each design, we add electrical nets to the active components and connect their pins to the chip’s peripheral I/O pads to enable active component control. 
We summarize the benchmark statistics in Table~\ref{tab:bench_stat}.
For the photonic computing tensor cores, Clements~\cite{NP_NATURE2017_Shen} and ADEPT~\cite{NP_DAC2022_Gu} circuits, we place the I/O pads along the top and bottom edges of the chip, while optical grating couplers are placed along the left and right edges. 
Compared to Clements, ADEPT contains more photonic components and nets. 
For the optical interconnect switches, GWOR~\cite{PD_tan2011generic}, both I/O pads and grating couplers are distributed along all four sides of the chip, with sufficient clearance reserved for packaging and coupling alignment. All benchmark circuits contain three available metal layers and one silicon waveguide routing layer.

\noindent\underline{\textbf{Baselines.}}~
Since no automated electrical router for PICs is currently available, we compare against an existing analog router \texttt{Anaroute}~\cite{PD_xu2019magical}, which supports multiple metal layers and arbitrary routing directions within each layer, thereby covering the basic requirements of PIC electrical routing. 
Based on this, we further construct a \emph{photonic-aware} variant as a second baseline (\texttt{Anaroute}$^\ast$): during path searching, we assign penalties to grid cells that overlap with photonic components, discouraging routing on top of optical components. 
For both baselines, we use the \emph{same netlist} produced after our escape routing and pad assignment. This isolates the routing algorithms from confounding factors in pad assignment; otherwise, a naive pad assignment can easily introduce unavoidable crossings and unfairly penalize the baselines.

\vspace{-7pt}
\subsection{PIC Electrical Routing Evaluation Results}
\noindent\underline{\textbf{Evaluation Metrics.}}~
Since our contributions focus on electrical routing for active PICs, we evaluate the quality of the metal routing solutions using the following metrics:
\textbf{(1)~Metal Layer Usage}: total number of metal layers used in routing, reflecting the planar routing capability of our routers, critical for PICs with limited metal layer stacks.
\textbf{(2)~Via Usage}: total number of vias, which impact signal integrity, parasitics, and reliability.
\textbf{(3)~User-Specified Rule Violations (USVs)}: designer-defined constraints beyond PDK rules, including: 
(a) Enlarged wire spacing requirements (e.g.,$0.5\mu m \rightarrow 3\mu m$); 
(b) Metal traces should try to avoid crossing over photonic components;
(c) Metal wires at M1 should avoid long parallel overlap with waveguides ($>20\mu m$ overlapping length counts as a violation);
(d) Metal wires at different layers should also avoid significant overlapping due to crosstalk concerns.
\emph{All DRC and USV checks are implemented using KLayout's DRC tool}.
\textbf{(4)~Runtime and Wirelength}:
We report the total routing runtime and electrical wirelength. 
For the low-speed PIC benchmarks considered (<1 GHz), electrical nets connecting I/O pads to pins primarily carry DC bias to program active photonic devices rather than high-speed signals as in VLSI sequential logic or RF ICs, so \emph{timing is not a critical metric for active low-speed PIC metal routing}.
\textbf{Wirelength serves only as an indirect hint of routability, rather than a measure of circuit performance}. 
Differences in wirelength shown later do not imply differences in PIC functionality or performance; therefore, it is not a focused metric.

\begin{figure}[t]
  \centering
  \includegraphics[width=0.99\columnwidth]{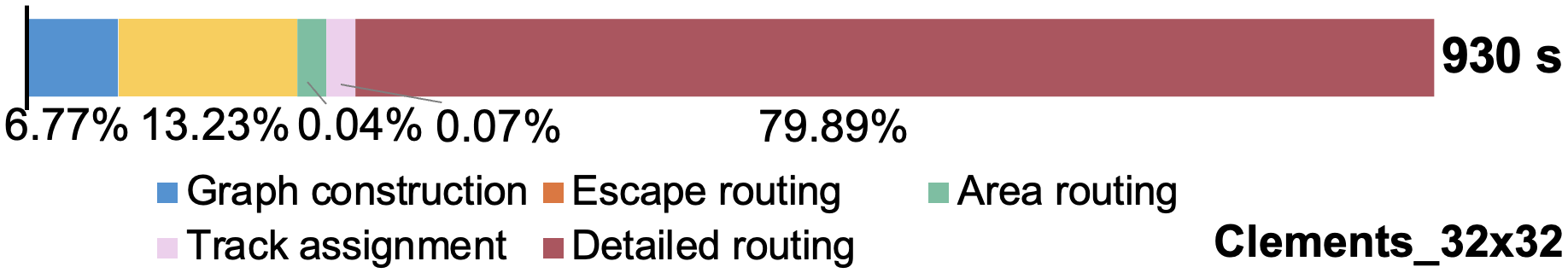}
  \vspace{-8pt}
  \caption{Runtime breakdown of our router.}
  \label{fig:runtime}
  \vspace{-5pt}
\end{figure}

\noindent\underline{\textbf{Main Results.}}~
Table~\ref{tab:result} reports quantitative comparisons of the routing results, and Fig.~\ref{fig:Layout} visualizes the layouts produced by our router. 
With crossing-aware global routing, our approach consistently requires fewer metal layers
and significantly fewer vias, while better satisfying designer constraints. 
Compared to the naïve baseline, we achieve a \textbf{$\sim$99\%} reduction in via usage, 
a \textbf{$\sim$98\%} decrease in \#USVs, 
and a \textbf{17$\times$} runtime reduction, yielding high-quality electrical routing under tight turnaround. 
In particular, on \texttt{Clements} and \texttt{ADEPT}, despite the large number of active devices,
(1) We obtain \textbf{planar} solutions that complete electrical routing using a single metal layer (M1).
For \texttt{GWOR}, pins must escape to the periphery, which increases the likelihood of crossings during escape routing.
Our router uses three metal layers, yet still requires far fewer vias than baselines.
This indicates that only a small fraction of nets climb to upper layers, and the majority remain on M1, yielding a \emph{near-planar} solution; the few nets that use vias can be manually adjusted if a planar layout is enforced. 
(2) A substantial portion of the \textbf{speedup} comes from our routing \emph{guidance}, which restricts detailed routing to narrow corridors around pre-prepared guidelines and thus greatly shrinks the search space. 
With guidance, we employ a fine grid step of 1 $\mu m$; the naive baseline must relax to 2 $\mu m$ to keep runtime manageable, as nets may exceed $1 mm$ in length, otherwise its runtime grows substantially. 
We also compare to manual routing in Fig.~\ref{fig:Manual}. 
Our router achieves comparable layout quality with orders-of-magnitude speedup (1h\(\rightarrow\)8s). 
Figure~\ref{fig:runtime} shows the runtime breakdown of our router, where detailed routing dominates the total runtime.
(3) Thanks to our \emph{track assignment}, we can \textbf{meet designer requirements} in global routing by enlarging inter-segment spacing and proactively avoiding inter-layer overlaps, thereby reducing post detailed routing USVs. 
In contrast, the baseline lacks spacing control and routes each net independently; it honors the basic design rules but cannot satisfy higher-level designer preferences. 
Note that our results incur only a small increase in wirelength, which is not the primary objective in this context.

\vspace{-5pt}
\subsection{Robust Support of Different Wire Pitch} 
\begin{table}[]
\caption{Our router can support high-quality electrical routing for \texttt{Clements\_8$\times$8} across different metal pitch settings.}
\vspace{-10pt}
\resizebox{\columnwidth}{!}{
\begin{tabular}{|c|c|c|c|c|}
\hline
Metrics & pitch=5$\mu m$ & pitch=10$\mu m$ & pitch=15$\mu m$ & pitch=20$\mu m$ \\ \hline
\#Via  & 0       & 0        & 0        & 2        \\
\#Layer & 1       & 1        & 2        & 3        \\
\#USV   & 2       & 1        & 4        & 6        \\
Runtime (s) & 33      & 36       & 35       & 38       \\
WL ($mm$) & 46.94   & 47.87    & 48.56    & 50.01    \\ \hline
\end{tabular}
}
\label{tab:pitch}
\vspace{-10pt}
\end{table}

Since our framework provides fine-grained control over wire spacing, we can evaluate robustness under different \emph{wire pitch} requirements. 
We sweep the spacing
constraints and update the corresponding \texttt{KLayout} DRC rules (see Table~\ref{tab:pitch}).
Across all settings, our router successfully honors the specified spacing requirements while maintaining low \#USVs and stable runtime. 
A larger wire pitch reduces the number of available tracks in each routing channel formed by photonic components, which forces more nets onto additional metal layers. 
As a result, the overall metal layer usage increases. 
Note that the reported \#Vias remains 0 in some cases because layer transitions inserted directly at pins/pads to move a net onto another metal layer are not counted; we only count vias introduced along routed paths during the detailed routing stage.

\vspace{-8pt}
\subsection{Effectiveness of Soft Guidance Cost} 
\begin{table}[]
\caption{\textbf{\emph{Soft}} guidance outperforms hard guidance with significantly higher routing quality and faster runtime.}
\vspace{-10pt}
\resizebox{0.9\columnwidth}{!}{
\begin{tabular}{|c|cc|cc|cc|}
\hline
\multirow{2}{*}{Metrics} & \multicolumn{2}{c|}{Clement\_8$\times$8} & \multicolumn{2}{c|}{ADEPT\_8$\times$8} & \multicolumn{2}{c|}{GWOR\_8$\times$8} \\ \cline{2-7}
                         & Hard       & \cellcolor[HTML]{D9D9D9}\emph{\textbf{Soft}}      & Hard      & \cellcolor[HTML]{D9D9D9}\emph{\textbf{Soft}}     & Hard     & \cellcolor[HTML]{D9D9D9}\emph{\textbf{Soft}}     \\ \hline
\#Via                   & 0               & \cellcolor[HTML]{D9D9D9}0              & 47             & \cellcolor[HTML]{D9D9D9}0             & 102           & \cellcolor[HTML]{D9D9D9}6             \\
\#Layer                  & 1               & \cellcolor[HTML]{D9D9D9}1              & 2              & \cellcolor[HTML]{D9D9D9}1             & 3             & \cellcolor[HTML]{D9D9D9}3             \\
\#USV                    & 2               & \cellcolor[HTML]{D9D9D9}2              & 36             & \cellcolor[HTML]{D9D9D9}2             & 179           & \cellcolor[HTML]{D9D9D9}19            \\
Runtime (s)                  & 37              & \cellcolor[HTML]{D9D9D9}33             & 923            & \cellcolor[HTML]{D9D9D9}65            & 1852          & \cellcolor[HTML]{D9D9D9}99            \\
WL ($mm$)                  & 46.84           & \cellcolor[HTML]{D9D9D9}46.94          & 97.75          & \cellcolor[HTML]{D9D9D9}100.70         & 183.20         & \cellcolor[HTML]{D9D9D9}184.84       \\ \hline
\end{tabular}
}
\label{tab:guide}
\vspace{-10pt}
\end{table}

A key feature of our framework is the \emph{soft guidance cost}, which endows detailed routing with error-correction capability. To validate this property, we compare against a \emph{hard guidance} baseline, i.e., we only raise the priority of routing grids that touch the guidance. 
From Table~\ref{tab:guide}, the guidance mechanism has a substantial impact on final routing quality. 
The \emph{hard guidance} cost fails to fully leverage the global-routing guidelines and often degenerates into a naive detailed router. 
There are two main reasons:
(1)~\textbf{Grid-step sensitivity.} Our guidelines are fine-grained. If the search frontier does not touch guideline cells (because the guidance may not be the shortest path, to avoid components or vias), the hard-boost rule may never be triggered, causing the search to miss guidance entirely.
(2)~\textbf{Imperfect/Conflicting guidelines.} Guidelines are not perfect due to pin-location constraints and local interactions; when a guideline collides with already routed nets, the hard-boost quickly becomes ineffective and can even hurt quality. On ADEPT and GWOR, we observe more crossings and longer runtimes with hard guidance. This is because, once the search deviates from the guidance and reconnects through a cross-layer neighbor, additional vias are introduced. The deviation also disrupts search continuity, leading to unnecessary expansions. In contrast, soft guidance provides "error-correction": it steers the A$^\ast$ search back onto, or near, the guidance corridor and then proceeds toward the target node.

\vspace{-5pt}
\section{Conclusion}
We introduce the first photonics-aware electrical router for large-scale active photonic integrated circuits, completing the missing piece of the PIC physical design automation flow and enabling the first truly end-to-end automated layout solution.
Our framework integrates novel photonics-aware global routing, sequence-consistent track assignment, and soft guidance–assisted detailed routing to produce a high-quality PIC layout with minimal layer/via usage while flexibly honoring designer preferences.
Crucially, we show that PIC electrical routing is not a simple extension of VLSI routing but requires new strategies that respect unique electronic–photonic interactions. 
Our customized router achieves superior layout quality, scalability, and up to 17$\times$ speedup over existing EDA tools on large-scale PIC benchmarks. 
This work lays the foundation for scalable, fabrication-ready EPDA, accelerating EPIC design cycles and paving the way for the next generation of manufacturable, energy-efficient, and high-performance electronic–photonic systems.

\end{document}